\documentclass[useAMS,usenatbib,usegraphicx]{mn2e}

\title[The silicate profile towards NGC4418. ] {The silicate absorption profile in the ISM towards the  heavily obscured nucleus of NGC 4418}
\author[P. F. Roche, et al ]{  P. F. Roche$^{1}$\thanks{E-mail: p.roche1@physics.ox.ac.uk (PFR)}, A. Alonso-Herrero $^{2}$,  O. Gonzalez-Martin $^{3}$   \\
$^{1}${Astrophysics, Department of Physics, University of Oxford, DWB, Keble Road, Oxford OX1 3RH} \\
$^{2}$ Instituto de Fisica de Cantabria, CSIC-Universidad de Cantabria, E-39005 Santander, Spain\\
$^{3}$ Centro de Radioastronom\'ia y Astrof\'isica (CRyA-UNAM), 3-72 (Xangari), 8701, Morelia, Mexico 
}

\begin{document}

\date{Accepted 2015 March 2.  Received 2015 February 24; in original form 2014 December 1}

\pagerange{\pageref{firstpage}--\pageref{lastpage}} \pubyear{2014}

\maketitle

\label{firstpage}

\begin{abstract}
The 9.7-$\mu$m silicate absorption profile in the interstellar medium provides important information on the physical and chemical composition of interstellar dust grains.  Measurements in the Milky Way have shown that the profile in the diffuse interstellar medium is very similar to the amorphous silicate profiles found in circumstellar dust shells around late M stars, and narrower than the silicate profile in denser star-forming regions.  Here, we investigate the silicate absorption profile towards the very heavily obscured nucleus of NGC 4418, the galaxy with the deepest known silicate absorption feature, and compare it to the profiles seen in the Milky Way.  Comparison between the 8-13~$\mu$m spectrum obtained with TReCS on Gemini and the larger aperture spectrum obtained from the Spitzer archive indicates that the former isolates the  nuclear emission, while Spitzer detects  low surface brightness circumnuclear diffuse emission in addition.  The silicate absorption profile towards the nucleus is very similar to that in the diffuse ISM in the Milky Way with no evidence of spectral structure from crystalline silicates or silicon carbide grains. 

\end{abstract}


\section{Introduction}

The profile of the silicate absorption band in the interstellar medium (ISM) has been investigated by a number of authors over the last 30 years.  Roche \& Aitken (1884, 1985) measured the 8-13$\mu$m spectrum of a number of reddened, dusty Wolf-Rayet stars and sources in the central parsec of the galaxy and concluded that the silicate absorption profile in the diffuse ISM is remarkably similar to the circumstellar silicate emission profile seen in late type giants. They adopted the silicate emission profile of the circumstellar shell of the supergiant $\mu$~Cephei as representative of the diffuse ISM.   The broad, smooth  profile suggests that the grains in the ISM are predominantly amorphous.  More recent studies include those of \citealt{Bowey98}, Chiar \& Tielans 2006 and van Bremen et al 2011, who came to similar conclusions.  These investigations have been used to place limits on the fraction of crystalline  silicates and silicon carbide in the ISM through the measured  limits on the sharp spectral structure that crystalline grains would produce (e.g. \citealt{Kemper04}, \citealt{Whittet90}).  

A substantial source of uncertainty in defining the silicate absorption profile in the ISM is that  the detailed spectral shape of the underlying emission is generally not known, but as the absorption optical depth increases, this effect become less important.  However,  uncertainties in the  flux level of the underlying emission, together with  calibration uncertainties, make it difficult to estimate the amount of grey extinction.  Fritz et al ( 2011) have used observations of infrared recombination lines towards the Galactic Centre to estimate the extinction at a range of near- and mid- infrared wavelengths and thus place some constraints on the relative extinctions in the continuum and the silicate absorption feature. 

In this paper, we present estimates of the silicate profile of the ISM in the galaxy NGC 4418 which has a compact,  very heavily obscured nucleus that is bright in the mid-infrared (Roche et al 1986).  

\section{Observations}

Observations of  NGC 4418 were obtained with the Gemini Thermal-Region Camera Spectrograph (T-ReCS) \citep{Telesco98} in June 2008 under Gemini programme GS-2008A-Q-55. The observations employed standard  chop-nod sequences, with a chop-throw of 15$^{\prime\prime}$.  The fixed pixel size of T-ReCS is 0.09$^{\prime\prime}$.     Spectra were obtained through a 0.35$^{\prime\prime}$ slit and the broad-$N$ filter, centred at 10.36-$\mu$m and with 50\% transmission between 7.70-12.97~$\mu$m.  The spectral resolution was 0.08$\mu$m, sampled at  0.022 $\mu$m/pixel.  The position angle of the slit was 30$^o$ and the total on-source  exposure time was 22 minutes.  Observing conditions were good, with relatively stable seeing conditions. 

Initial data reduction was conducted using the T-ReCS {\sc iraf} pipeline. The chop frames were mean-averaged and the alternating positions subtracted to remove residual background structure.  Examination of the acquisition image, which showed only an unresolved source, suggested there would be no significant off-source emission within the chop throw. The spectra were centred and straightened, though  the corrections applied were minimal.  

The extremely deep silicate absorption feature results in a dramatic reduction in flux between the relatively lightly obscured wings of the band near 8 and 12~ $\mu$m and the silicate minimum in the 9 -10~$\mu$m region.  Careful attention was paid to residual chop-nod emission and the effects of cross-talk between the different read-out channels on the detector.  Weak residual spectral structure above and below the rows containing flux from NGC 4418 was coadded and smoothed before subtracting from the array image.  The silicate absorption is so deep that the flux at the minimum is $>$200 times fainter than in the wings and is not detected right in the minimum near 9.7~$\mu$m.  Wavelength calibration was applied using the sky emission line list supplied by Gemini for the $N$-band.  Similar processing was applied to the standard star spectrum of HR4432. Telluric-correction and flux-calibration were performed by dividing the averaged galaxy spectra by the standard star spectrum and multiplying by a 4100 K blackbody appropriate to  the spectral-type of K3.5{\sc iii} of  HR4432 (\citealt{Johnson66}). 
The resulting  spectrum is plotted in Fig. 1., where  the error bars are estimated from the scatter in the points.  Spectra were also extracted with 3-pixel bins to search for any wavelength-dependent spatial structure. 

The low-resolution (R$\sim$100) spectrum of NGC 4418 obtained with Spitzer  was downloaded from the enhanced data product archive at the Spitzer Science Center. The data were taken under Spitzer campaign ID IRSX005200 and the P.I. was J. Houck. The slit width of the low-resolution, short wavelength module  of the IRS  at 10~$\mu$m is 3.7 arcsec (Houck et al 2004).

\begin{figure}
\begin{center}
\includegraphics[width=8.8cm]{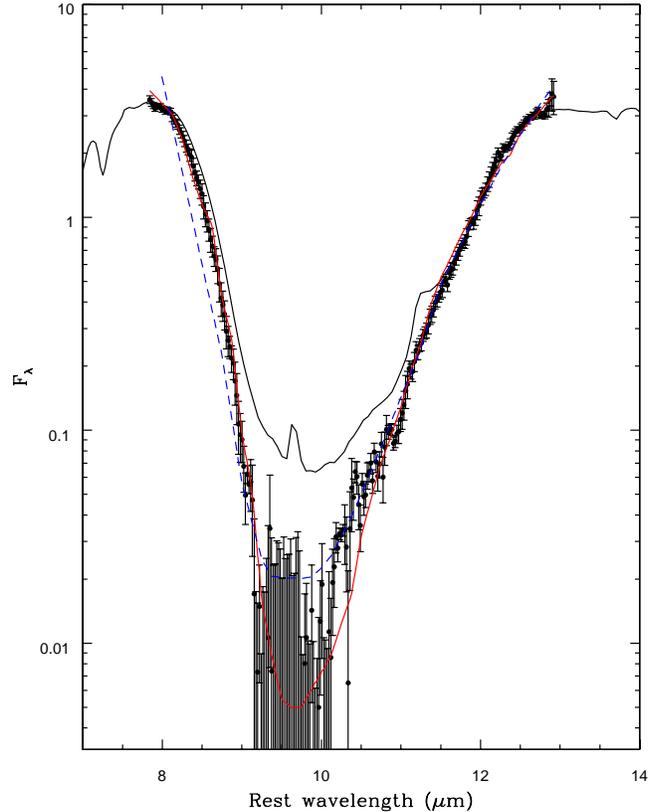}
\caption{The spectra of the nucleus of NGC 4418 obtained with Spitzer (solid line) and TReCS filled circles,  in units of 10$^{-18}$ W~cm$^{-2}$~$\mu{\rm m}^{-1}$. The thin solid (red) and dashed  (blue) lines represent fits to the TReCS spectrum using the $\mu$~Cep and Trapezium silicate profiles respectively. } 
\label{fig:n4418spec}
\end{center}
\end{figure}

\section{Results}

\subsection{Imaging}
The acquisition image obtained in the broad N-band filter shows a bright, compact mid-infrared object with no evidence of extended emission. The FWHM of the image is 3.7 x 3.7  pixels, corresponding to 0.33 arcsec. This compares with the measured FWHM of the comparison star of 3.7 x 3.5 pixels and a diffraction-limited width of 0.28 arcsec in the N-band filter.  From the similarity to the comparison star, we conclude that NGC 4418 is unresolved in the N band. This  agrees with the results of \citealt{Evans03} who also found NGC 4418 to be unresolved  between 8 and 24~$\mu$m. The good agreement in image size between the galaxy and the star implies that the bulk of the flux must be contained within  a region smaller than the diffraction-limit; subtraction in quadrature of the 0.28" diffraction-limited size from the measured FWHM suggests a size $<$0.2 arcsec for the mid-infrared emission from NGC 4418.  At a distance of $\sim$34Mpc (\citealt{Braatz97}), this implies a physical diameter of $<$33~pc.

\subsection{Spectroscopy}

The TReCS 8-13~$\mu$m spectrum (figure \ref{fig:n4418spec}) shows the deep silicate absorption feature seen in previous observations, with a reduction in flux of a factor of $>$100  between the wings of the silicate absorption band at 8 and 13~$\mu$m and the silicate minimum at 9.7~$\mu$m.  The spectrum obtained by Spitzer (\citealt{Spoon07}) is also plotted in Fig \ref{fig:n4418spec}. The calibration uncertainties suggest that the very close agreement between the two spectra may be partly fortuitous, but an independent reduction of the TreCS data using the RedCan pipeline by Gonzalez-Martin et al (2013) gives essentially the same result. It is clear that most of the flux detected in the large Spitzer aperture is admitted by the 0.35~arcsec TReCS slit, confirming that the emission is very compact. However, there are clear differences in the spectral shape, with the S(3) H$_2$ emission line at 9.66~$\mu$m and weak 11.3~$\mu$m PAH emission evident in the Spitzer spectrum, together with a higher flux level in the silicate minimum. It is very likely that these spectral features  arise from very low level diffuse emission around the compact nucleus, and suffer substantially lower levels of extinction than the compact core.  Imanishi et al (2004) have detected prominent H$_2$ emission in their 2~$\mu$m spectrum together with the 3.3~$\mu$m PAH emission band within the central 2-4 arcsec. By contrast, the TReCS spectrum shows no evidence for any spectral structure other than the prominent silicate absorption.

Spatial cuts across  the spectrum of  NGC 4418 yield  FWHM increasing from ~3.5 to 4.3 pixels from 8 to 13~$\mu$m, consistent with an increasing contribution from diffraction with wavelength.  This behaviour is mirrored in the width of the spectrum of HR4432.  Examination of the width of the NGC 4418 spectrum showed no evidence for a significant change in width near the silicate minimum between 9 and 10.5~$\mu$m, although the low flux in this region means that the limits are not very stringent.   This suggests that we are not looking into a resolved optically thick surface at any wavelength  between 8 and 13~$\mu$m, but that the mid-infrared source is very compact across the N-band.  

The spectrum can be fit adequately using grey body underlying emission suffering absorption by a layer of cool silicate grains.  The fits with the narrow $\mu$ Cephei silicate profile, taken to be representative of the diffuse ISM, are substantially better than those using the Trapezium profile, with values of $\chi^2$/N of 2.4 and 6.5 respectively.  Even the $\chi^2$ value obtained with the $\mu$~Cep profile indicates a rather poor fit reflecting the increasing departures from the silicate profile near the absorption minimum, but it is a much better match than the Trapezium curve. The Trapezium  profile, which is extracted from emission from warm silicates around Theta 1 Ori (Gillet et al. 1975)   is taken to be representative of silicate emission from molecular clouds (Roche \& Aitken 1985).   It is wider than the silicate absorption in NGC~4418, and cannot provide a good fit to the short wavelength  edge of the band (see. Fig. \ref{fig:n4418spec}).  This is in line with results found in other nuclei with deep silicate absorption bands where the silicate profile is similar to that found in the diffuse ISM in the Milky Way (Roche et al 2007). The absorption optical depth given by the $\mu$ Cep  fit is $\tau_{9.7\mu{\rm m}} =7$, in agreement with earlier observations (Roche et al 1986).   The apparent depth of the silicate absorption in the Spitzer spectrum is significantly lower, but it appears that this is a result of the inclusion of additional extra-nuclear emission in the larger entrance aperture of the IRS.  Fits to the Spitzer spectrum with either silicate curve give  absorption optical depths of $\tau_{9.7 \mu{\rm{m}}}\sim$ 4 - 4.5, but the fits are very poor.   The spectral structure near 11.2~$\mu$m indicates that PAH emission contributes significantly within the large Spitzer aperture and the lack of sharp structure near the 12.7~$\mu$m PAH band (Roche, Aitken \& Smith 1989) indicates that the PAH emitting region suffers much lower extinction than the continuum at 10~$\mu$m.

\section{Silicate Profile}

Detailed inspection of the fits to the T-ReCS spectrum shows that the $\mu$ Cep profile matches the wings of the silicate profile between 8 and 9~$\mu$m  and 11 and 13~$\mu$m very well, but that it starts to deviate in the silicate minimum between 9 and 11~$\mu$m, falling below the measured flux.  Several possibilities for this discrepancy suggest themselves:  the $\mu$ Cep profile may be too strongly peaked in this region to fit the observed spectrum; there may be a contribution from diffuse emission in NGC 4418 that partially fills in the silicate minimum; there could still be residual uncorrected thermal emission from the telescope or detector crosstalk effects that affect the zero level and hence the measured fluxes at such low levels.   However, comparison between the Spitzer and TReCS spectra suggests that diffuse emission produces the S(3) H$_2$ line, and the 11.3$\mu$m PAH feature and that the associated continuum fills in the silicate minimum; it is  probable that this component contributes low level flux   at 9 - 11~$\mu$m in the TReCS spectrum.

\begin{figure}
\includegraphics[width=6.6cm, angle =-90]{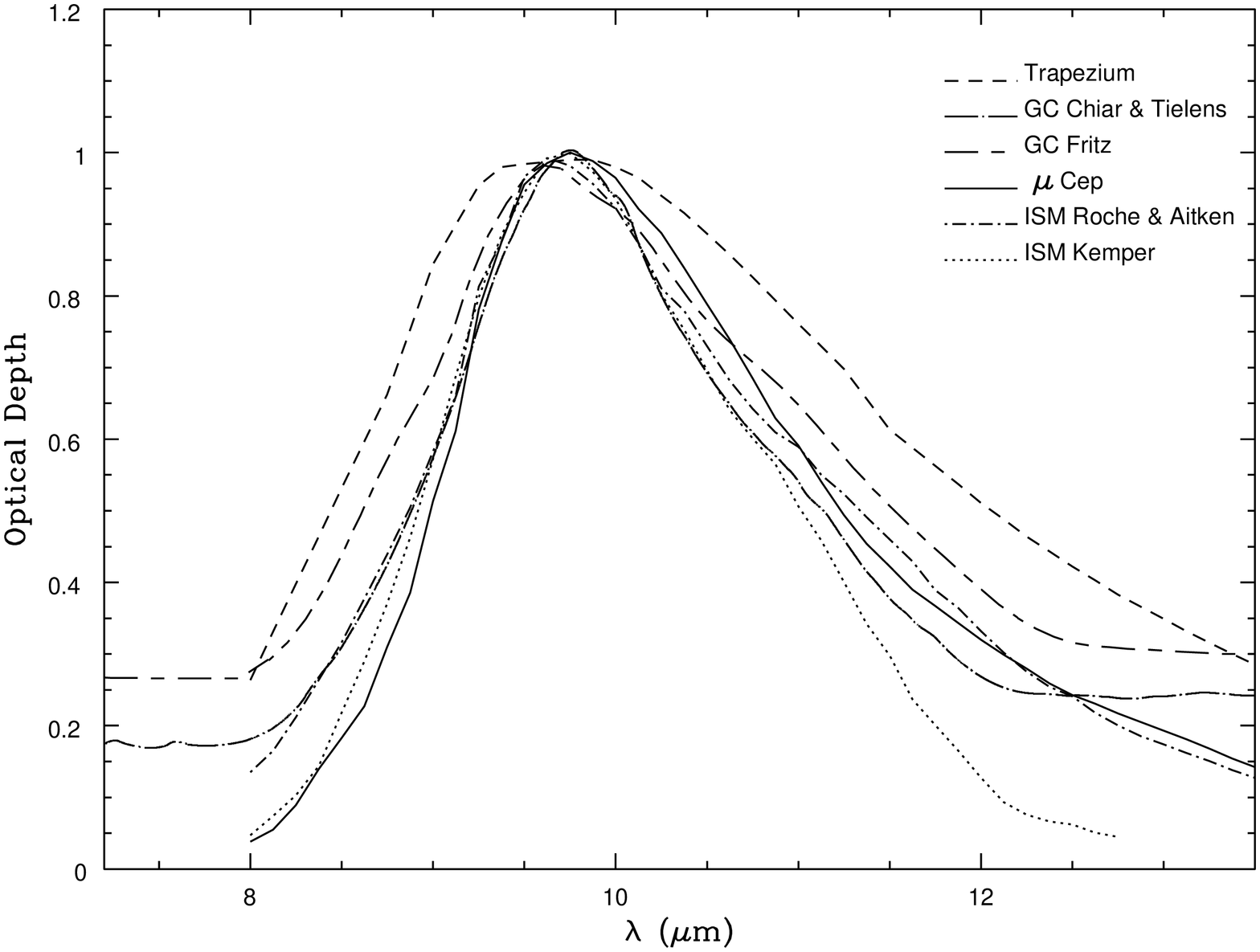}
\includegraphics[width=6.6cm, angle =-90]{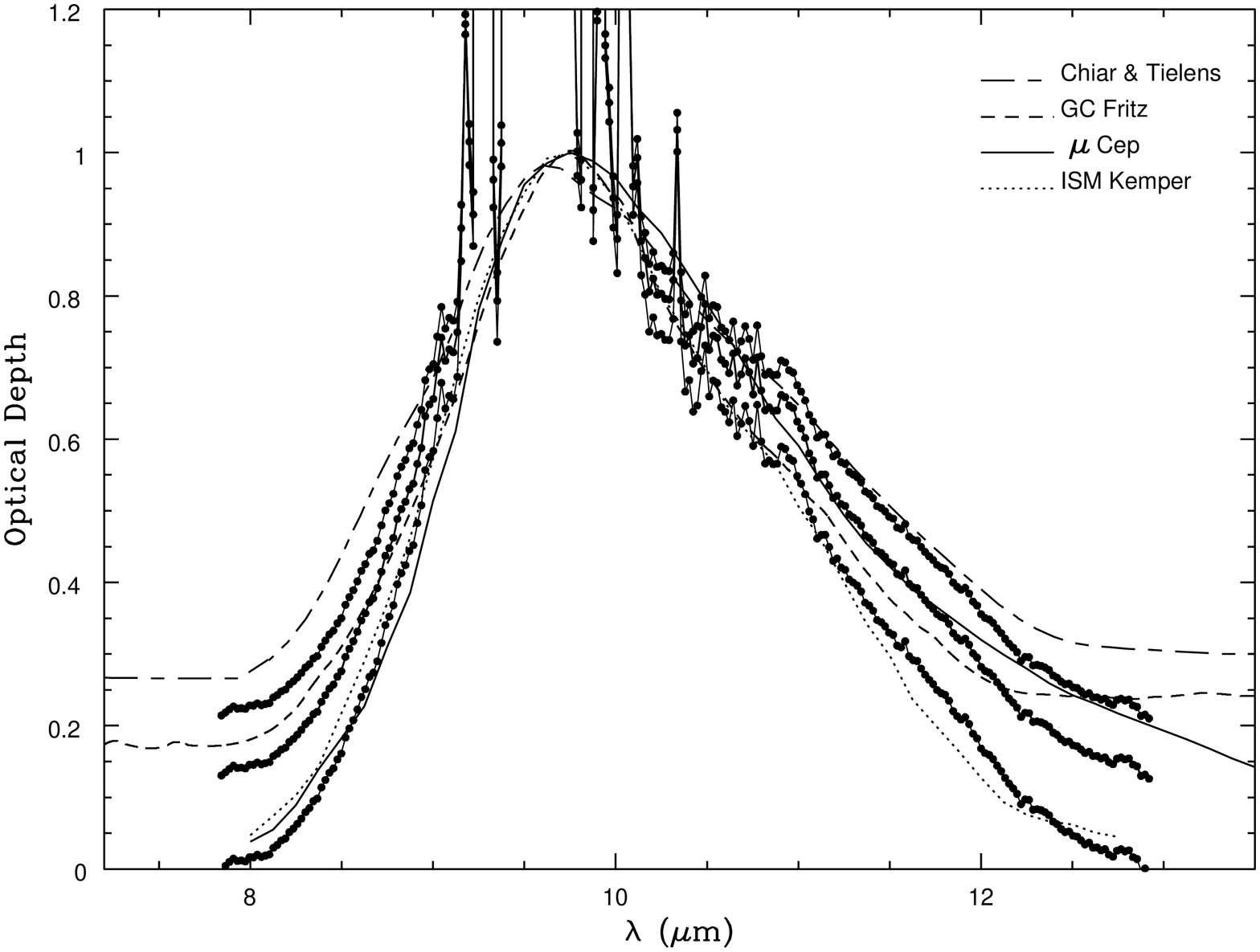}
\includegraphics[width=6.6cm, angle =-90]{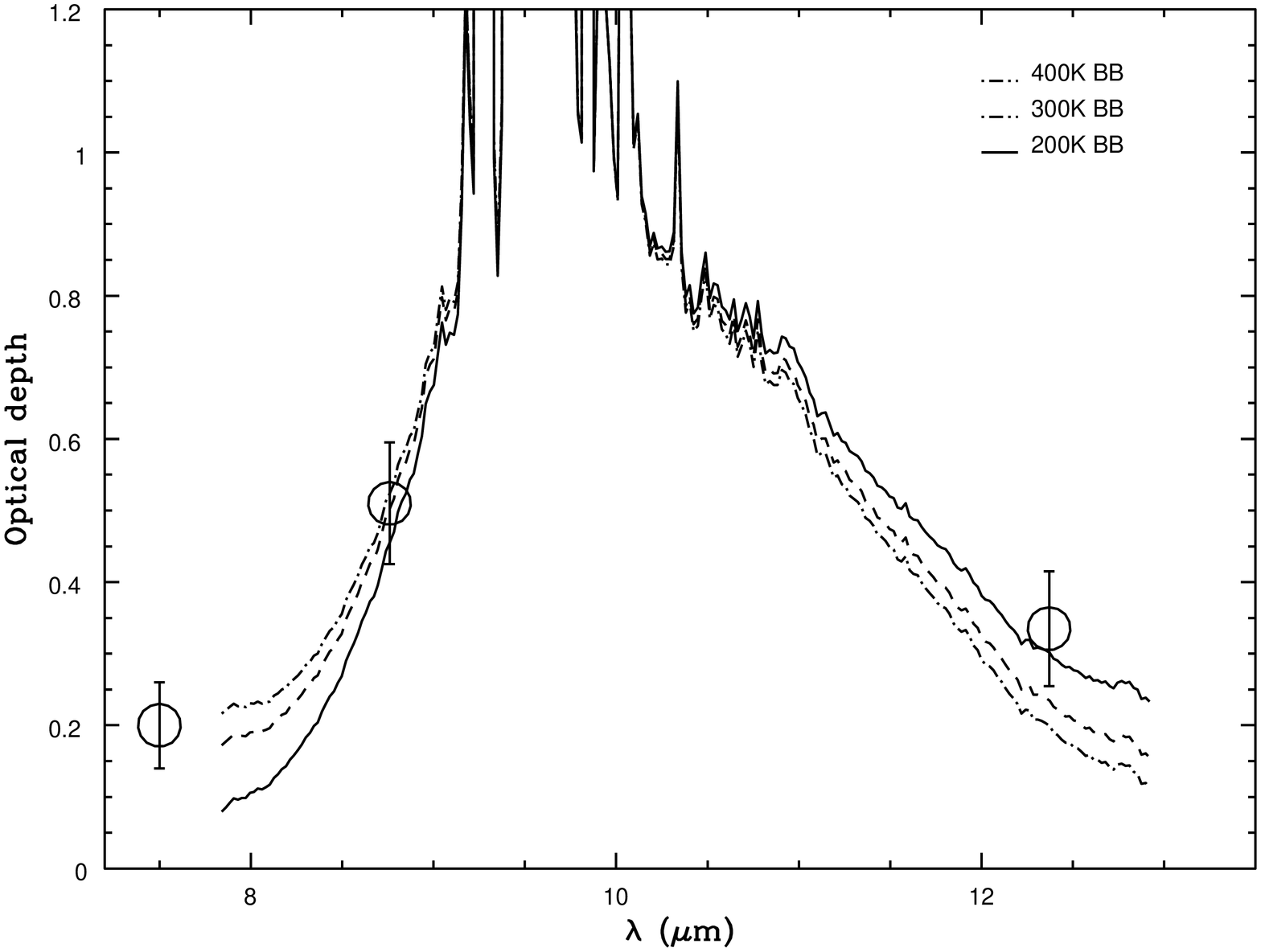}
\caption{Normalised optical depth profiles from the literature and derived from the TReCS spectrum of NGC 4418.  (a) Top:  silicate profiles from published papers (see text).  (b)  Middle: NGC 4418 opacity profiles obtained from power-law emission with varying adopted flux levels compared to the profiles derived from $\mu$~Cep and ISM profiles from  Roche \& Aitken (1985), Fritz et al (2011), Chiar \& Tielens (2006)  and  Kemper et al (2004).  (c)  Bottom: NGC 4418 opacity profiles obtained from 200K, 300K and 400K underlying blackbody functions.  The large open circles represent the opacities determined from hydrogen recombination lines by  Fritz et al (2011). } 
\label{fig:profs}
\end{figure}

To investigate the silicate profile further, we adopt models for the underlying emission from the nucleus of NGC 4418 and examine the absorption profiles required by the observed spectrum. There is no evidence of sharp spectral structure that could arise from PAH or line emission in the TReCS spectrum of NGC 4418.  We therefore investigated black-bodies, power-laws and a combination of these with silicate emission as possible emitting sources.  We used a blackbody fitted to the Spitzer IRS spectrum at 5.5 and 15~$\mu$m to provide estimates of the minimum level of the continuum emission, and then varied the level of underlying emission to investigate the range of optical depth profiles generated by the observed TReCS spectrum.  Examples of optical depth curves generated from underlying power law and blackbody  emission spectra  are shown in figure \ref{fig:profs} together with curves from the literature.  

Varying the spectral slope of the power laws or the temperature of the underlying blackbody emission has only a very small effect on the silicate profile, but does affect the slope of the associated continuum; this is illustrated in Fig 2c where the profiles generated from adopted underlying blackbody emission at 200, 300 and 400~K, and normalised at 10~$\mu$m are plotted.   Adjustment of the absolute level of the underlying emission affects the level of the continuum in the silicate profile.  The minimum level of the continuum is well-defined and is given by setting the opacity of the silicate feature to fall to zero at the extremes of the spectrum at 7.6 and 13~$\mu$m.  This curve is shown in Fig. 2b along with others with higher continuum levels.

The low signal-to-noise ratio of the spectrum in the silicate minimum between 9 and 10.5~$\mu$m leads to a large scatter in the optical depth curves in this wavelength range,  but the profiles are well-defined outside this region.  As anticipated from the fits described above, the Trapezium profile is substantially broader than any of the optical depth curves produced, and no combination of underlying emission spectra can produce a satisfactory match with the Trapezium absorption profile. Rather, the optical depth profile towards NGC 4418 is a much closer match to the diffuse ISM profiles suggested by Roche \& Aitken (1984), Chiar \& Tielens (2006)   and Kemper et al (2004).   The effects of varying the level of the underlying emission are illustrated by the grey solid lines in figure \ref{fig:profs}b.  The minimum level assumes no grey extinction and provides an optical depth curve for the extinction relative to that at 7.8 and 13~$\mu$m.  This is similar to the silicate profiles generated by some other workers, e.g. Kemper et al (2004) who adopted continua near 7.5 and 12.3~$\mu$m to estimate the silicate profile within these limits.   The close match between the optical depth profile derived from NGC 4418 and the ISM profiles confirms the similarity between the dust along the line of sight to the nucleus of NGC 4418 and that towards the centre of the Milky Way. 

Because the optical depth to the emitting regions of NGC 4418 is so high, the addition of silicate emission to the power-law and blackbody functions has very little effect on the derived opacity curves; essentially the profile is dominated by the absorption component.  This is borne out by the fits to the spectra, where the addition of silicate emission components does not significantly improve the values of $\chi^2$ obtained, though it does increase the  absorption optical depth. 

Fritz et al (2011) have estimated the extinction at a range of infrared wavelengths from observations of hydrogen recombination lines.  Their estimates of the relative extinctions can be used to place some constraints on the relative extinction levels at the wavelengths of the lines at 7.5, 8.76 and 12.37~$\mu$m.  Because the long-$\lambda$ recombination lines are relatively weak, the uncertainties on the measurements and hence of the extinction determinations are significant, but they do provide important guidance, indicating a significant continuum opacity at 7.5$\mu$m which will contribute at a level of $\sim 20\%$ of the peak at 9.7~$\mu$m.   These data are shown in Fig. 2c, together with curves that illustrate the effect of varying the temperature of an underlying blackbody continuum. 

 \begin{figure*}    
\begin{minipage}[t]{0.35\textwidth}
\includegraphics[width=\linewidth, angle =-90]{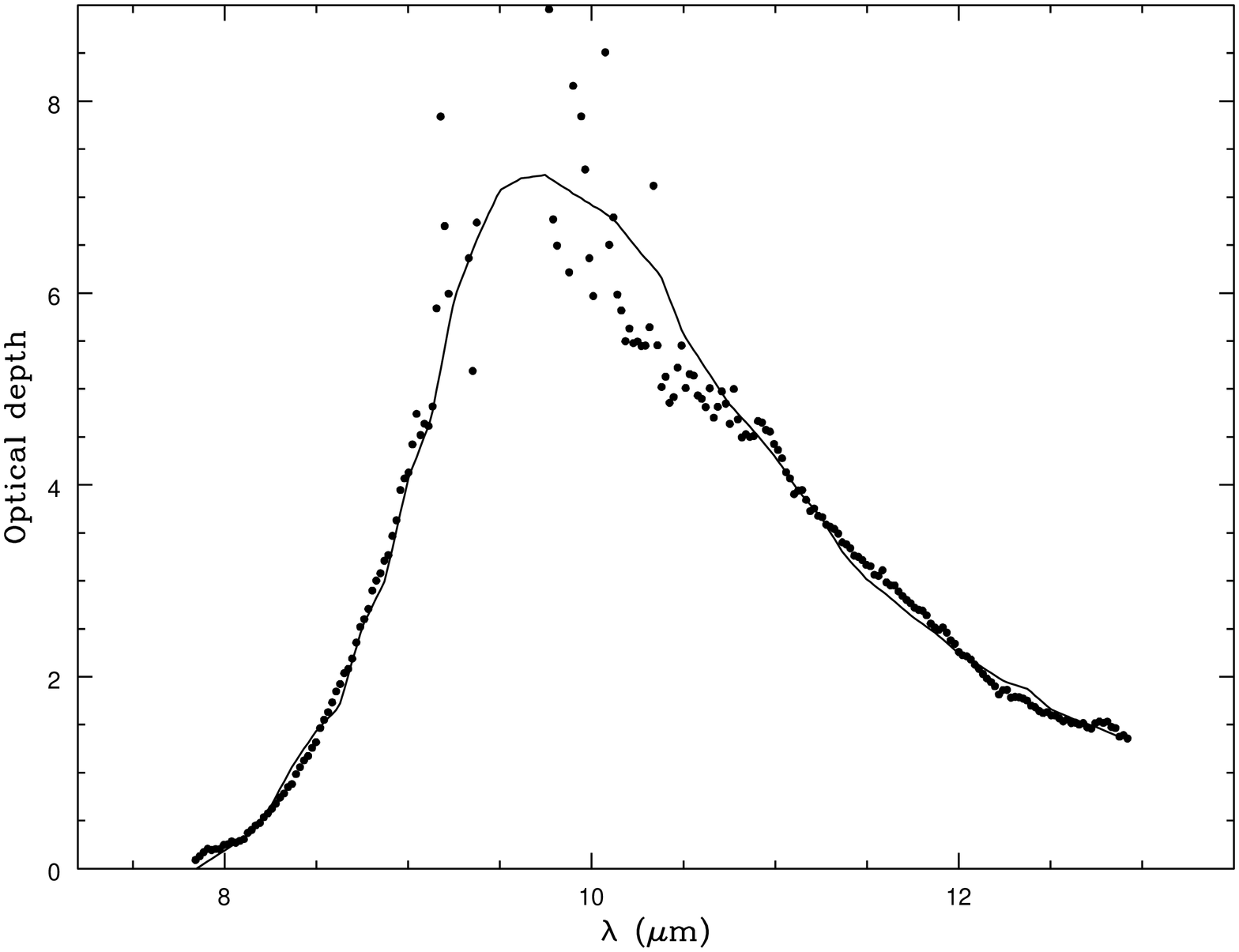}
\end{minipage}
\hspace{3cm}
\begin{minipage}[t]{0.35\textwidth}
\includegraphics[width=\linewidth, angle =-90]{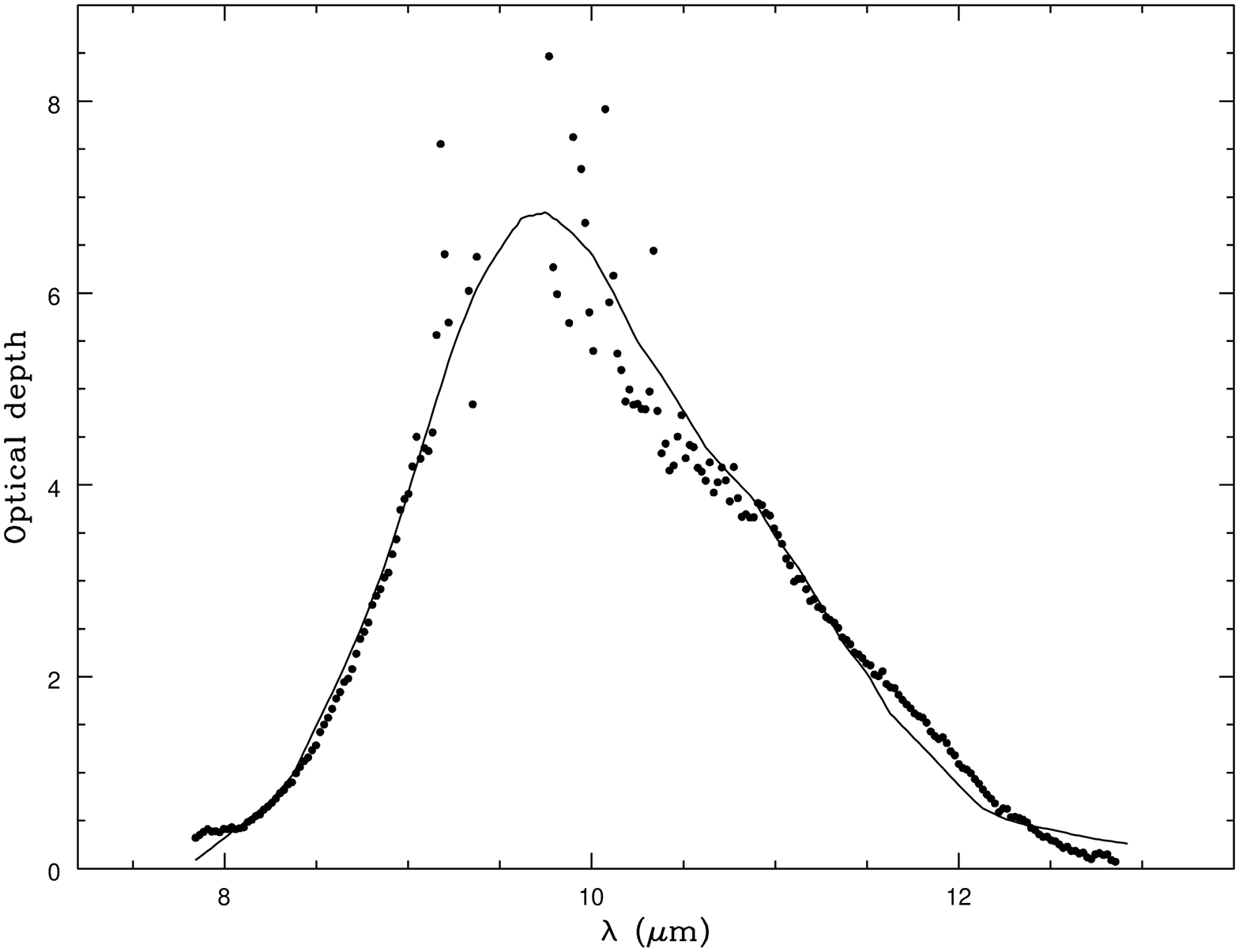}
\end{minipage}

\vspace*{0.1cm} 
\begin{minipage}[t]{0.35\textwidth}
\includegraphics[width=\linewidth,angle =-90]{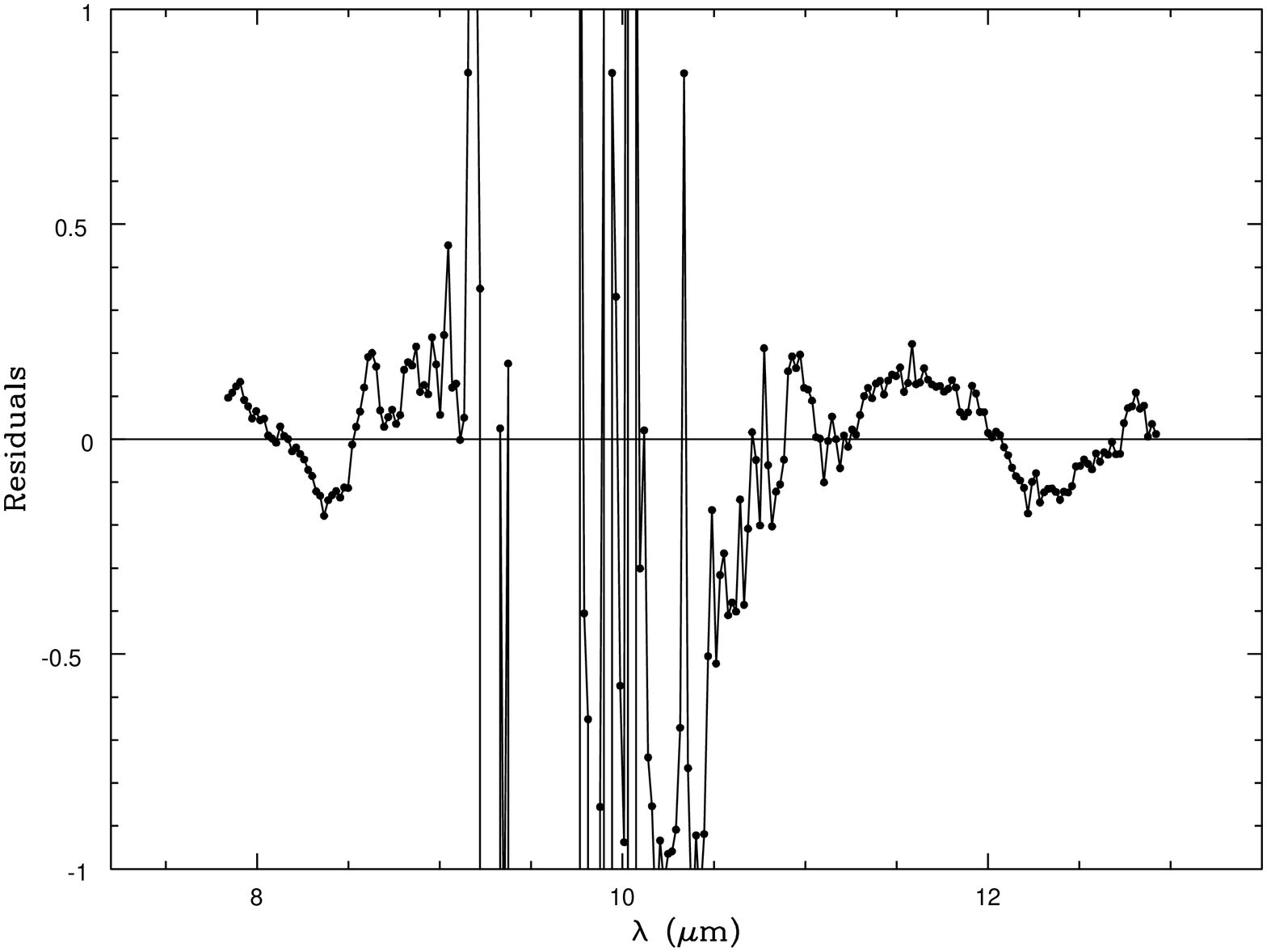}
\label{fig:distal}
\end{minipage}
\hspace{3cm}
\begin{minipage}[t]{0.35\textwidth}
\includegraphics[width=\linewidth, angle =-90]{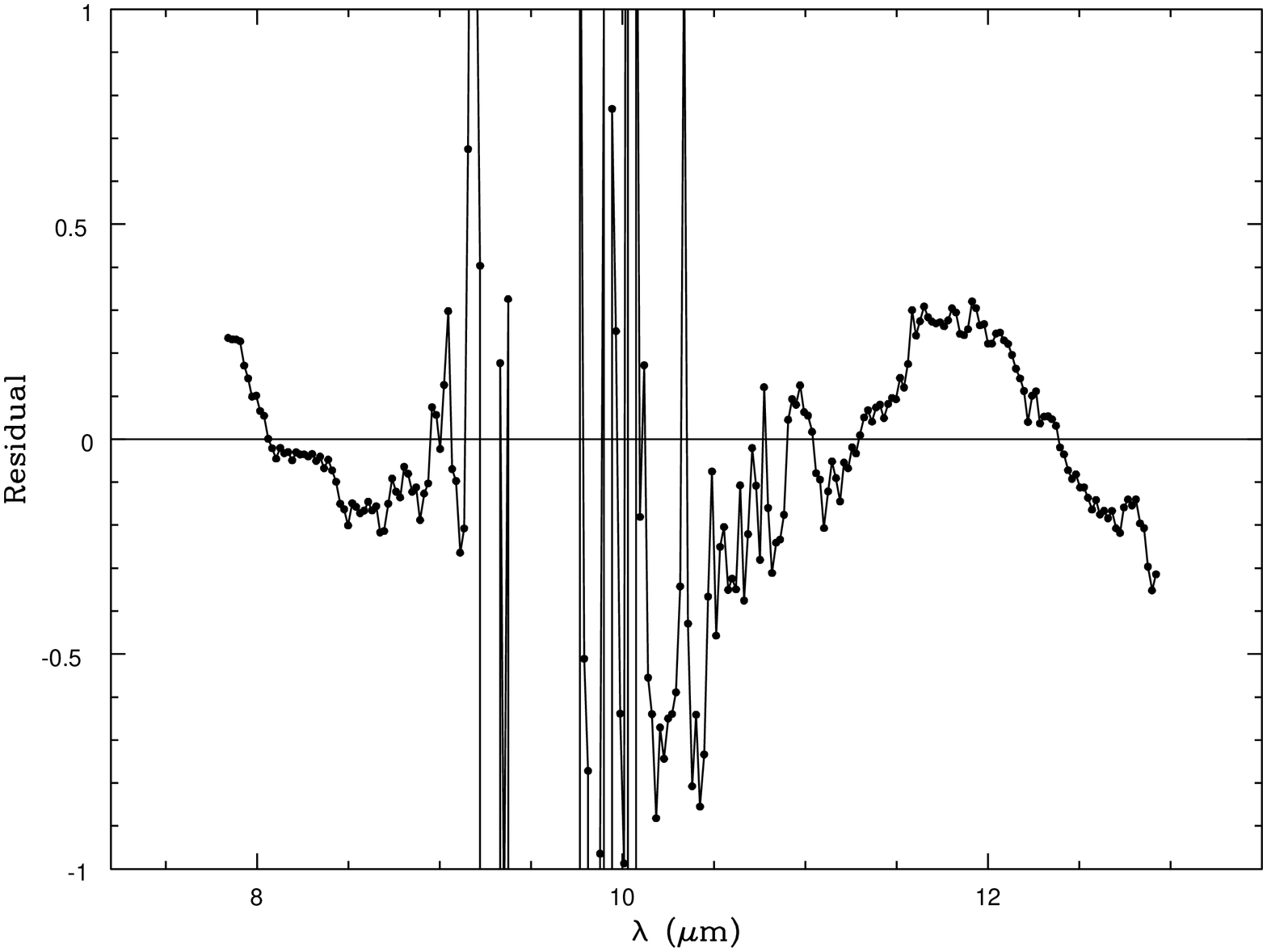}
\label{fig:combined}
\end{minipage}

\caption{Optical depth spectra of the nucleus of NGC 4418 obtained using the $\mu$ Cep emissivity curve (top left) and the silicate profile from \citet{Kemper04} (top right).  The residual structure between the observed spectrum and the fits are shown in the bottom plots and allow estimates of possible contributions from crystalline silicates or other dust  components. } 
\label{fig:dered}
\end{figure*}

\subsection{Contributions from Crystalline Silicates or Silicon Carbide?}
 There is no evidence of contributions from silicon carbide or crystalline silicates to the N-band NGC~4418 absorption profile, suggesting that in this galaxy, like the Milky Way, the silicates are predominantly amorphous.  Figure 3 show the optical depth profiles of the TReCS spectrum of NGC~4418 together with the best fit  $\mu$ Cephei and Kemper silicate opacity curves.  The lower plots show the  residual optical depths  i.e. the differences between the NGC 4418 optical depth curves and the fits using the $\mu$ Cep and Kemper curves.  In both cases, the residual optical depth is $\Delta(\tau_{\lambda}) < 0.2$ between 8 - 9~$\mu$m and 11-13~$\mu$m. The ratio of the maximum residual to the peak optical depth is similar to that found by Kemper et al (2004) using amorphous silicate curves to fit the Galactic Centre profile, indicating that the contributions from high band strength crystalline silicates in the N-band along the line of sight to the nucleus of NGC 4418 are similar to those found for the Milky Way.  Kemper et al (2005) place an upper limit on the crystalline fraction  of 2.2$\%$, but Li et al (2007) argue that this upper limit is increased to $\sim3 -5\%$ when the presence of icy mantles on the silicate cores is taken into account.

To quantify limits of any contributions from SiC or crystalline silicates, fits to the TReCS spectrum were conducted with a power-law or black-body emission functions suffering extinction by silicates and SiC. Laboratory studies have generated a range of crystalline silicate profiles, and although there are good matches to some of the features in circumstellar shells and disks  (see e.g.  \citet{Molster05}) these features have not yet been detected in the general ISM.  There are few good examples of crystalline silicate absorption profiles, but Spoon et al (2006) have detected a number of sharp absorption bands in Spitzer spectra of Ultra-Luminous Infrared Galaxies that they attribute to crystalline silicates. We have employed the  crystalline silicate emissivity function derived from IRAS 08572+3915 (\citet{Spoon06}) as the best estimate of a crystalline silicate component in the ISM. Including a component with this  profile in the fits to the NGC~4418 spectrum with the $\mu$~Cep profile does not reduce the value of $\chi^2$ and we conclude that the upper limit to the contribution from a crystalline silicate peaking at 11~$\mu$m is $\tau_{11\mu{\rm m}} < 0.1$ compared to $\tau_{9.7\mu{\rm m} }= 7.2$ for the $\mu$~Cep amorphous silicate component.  

The absence of crystalline silicate signatures in the N-band is surprising as Stierwalt et al (2014) have attributed an absorption band at 24~$\mu$m in the Spitzer spectrum of NGC 4418 (and several other heavily obscured luminous infrared galaxies) to crystalline silicates. Spoon et al (2006) found that all of the most heavily obscured ultra luminous infrared galaxies  in their sample displayed evidence of crystalline silicates, although only 08572+3915 showed a definite 11~$\mu$m crystalline silicate absorption feature.  

Similar fits were undertaken to provide estimates of any contribution from silicon carbide to the absorption in NGC~4418.  We used the emissivity profile derived from the carbon star Y Tau (see Aitken \& Roche 1982) as representative of SiC injected into the ISM.  Adding this component to the fit reduces the value of $\chi^2$, from 2.5 with silicate absorption alone to 2.25,  with an optical depth  $\tau_{11\mu{\rm m}}$= 0.15 at the peak of the SiC band.  Inspection of the fit shows that the introduction of the broad SiC component provides increased absorption beyond 11~$\mu$m which reduces  the optical depth of the silicate absorption component by 0.2. Because of the breadth of the SiC feature, it is difficult to judge whether this is a real feature or rather results from uncertainties in the detailed profile of the silicate absorption band. We therefore take the upper limit to the SiC absorption feature of $\tau_{11\mu{\rm m}}$ =0.3 as the limit on the optical depth of any SiC component.    This limit provides an estimate of the amount of silicon locked up in silicon carbide of $<4\%$ in the ISM in NGC~4418 which is comparable to the limits reported by \citet{Whittet90} along the line of sight to the Galactic Centre and by \citet{Chiar06} towards dusty Wolf-Rayet stars in the Milky Way.     

\section{The obscured nucleus of NGC 4418}

Many of the salient properties of the core of NGC 4418 are summarised by Sakamoto et al (2013). The deep silicate absorption with {$\tau_{9.7\mu{\rm m}} \sim 7$} corresponds to a visual extinction of $\sim$100 magnitudes for a cold foreground screen  or more if we are looking into an optically thick source. Other extinction estimates have been derived from x-ray and mm-wave observations which indicate a compact, dense warm core which may harbour a  Compton-thick AGN (e.g. Sakamoto et al 2013, Maiolino et al 2003), while the non-detection of NGC 4418 by SWIFT/BAT2 (Koss et al 2013) may  also be consistent with extremely high levels of obscuration.  Recent radio wavelength interferometry by Varenius et al (2014) has revealed a cluster of compact sources within the central 0.2 arcsec, which may arise from young star clusters and/or an AGN in the nuclear regions. One or more of these objects presumably emits the mid-IR flux, and they add to the picture of a very compact source or compact group of sources in the nuclear regions. 

The core is compact at all wavelengths and accounts for the bulk of the infrared output of 10$^{11}$ L$_\odot$, with the source of this emission lying behind a very high column density  of obscuring dust.  The very high column density and spectral energy distribution indicate that the dust emission from the core remains optically thick out to far-infrared wavelengths (Roche \& Chandler 1993, Sakamoto et al 2013).  The fact that the silicate absorption profile towards the core is narrow and consistent with the profiles found within the diffuse ISM in the Milky Way suggests that the bulk of the observed mid-infrared extinction arises in  diffuse material surrounding the dense molecular core.  This diffuse material must have a very high covering factor, at least from the viewpoint from the Earth.  

It has been argued that the dust along the line of sight to galaxies with very deep silicate absorption features is predominantly located in the extended structures of the host galaxy rather than in the central region or molecular torus in active nuclei.  This conclusion is supported by the narrow silicate profile of the most heavily extinguished objects (Roche et al 2007), the fact that AGN host galaxies viewed at high inclination have deeper silicate features than those at low inclination (Alonso-Herrero  at al., 2011),  and the poor correlation with silicate depth of Compton-thick AGN (Goulding et al., 2012, Gonzalez-Martin et al., 2013).  NGC~4418 may be an extreme example of this, with the bulk of the silicate absorption arising in the diffuse ISM, perhaps in the dusty structures seen in the NICMOS images (Evans et al., 2003)
 
Costagliola et al (2013) have estimated that cool material surrounding the compact nuclear core of NGC 4418 has an atomic HI  gas column density of at least $2 \times 10^{22}$~cm$^{-2}$. This is about  an order of magnitude lower than the hydrogen column density inferred from the $\sim$100 magnitudes of visual extinction corresponding to the silicate absorption feature, adopting a ratio of N({\sc h}) $= 2.2 \times 10^{21}$ A({\sc v}) cm$^{-2}$ (e.g. Guver \& Ozel 2009).  However, although the core is very compact at both IR and radio wavelengths, the underlying emission sampled by the silicate absorption may well have a different spatial extent to the radio continuum, and so we should perhaps not expect close agreement.  More puzzling are the unusual spectral properties of NGC~4418 in the 5-8~$\mu$m region where strong absorption features attributed to ice and hydrocarbons have been detected (Spoon et al 2001, 2005). The strong, broad absorption at 6~$\mu$m attributed to water ice, which has an optical depth $\tau_{6\mu{\rm m}} \sim 0.8$ (Spoon et al 2001),   strongly suggests that there is significant cold molecular material along the line of sight. It appears that the line of sight towards the nucleus of NGC~4418 incorporates some molecular material, but as is the case towards the Galactic Centre (e.g. \citealt{Chiar00}), the dominant component of silicates is representative of the diffuse ISM.  The ratio of optical depths in the 6~$\mu$m ice band to the silicate feature is about a factor of 3 higher towards NGC~4418 than towards Sgr~A, suggesting that the line of sight to the former passes through more cold dense material. 

Unfortunately, the non-detection of emission features in the mid-infrared spectra from the ground (and the absence of ionic emission lines in the Spitzer spectrum) means that the nature of the  source(s) emitting the mid-IR continuum remains unclear. It is however clear that we are not observing the spectrum of a typical galaxy dominated by  nuclear star formation (e.g. Roche et al 1991) but rather seeing into a region with bright, warm, compact and almost certainly optically-thick,  continuum emission through an absorbing layer of predominantly diffuse material.

\section{Conclusions}

The silicate absorption profile towards the nucleus of NGC~4418 is very similar to the absorption profiles measured in the diffuse ISM in the Milky Way, with no evidence of spectral structure that would arise from substantial contributions from  crystalline silicate or silicon carbide components. The profile is consistent with those produced towards obscured objects in the Galactic centre and because of its high optical depth, $\tau_{9.7\mu{\rm m}} =7$, the shape of the profile is relatively insensitive to the spectrum of the underlying emission component. 
 Limits on possible contributions by crystalline silicates or silicon carbide grains to the absorption profile towards NGC~4418 are similar to those estimated towards obscured objects in the Milky Way.
 
It is likely that much of the silicate absorption at mid-infrared wavelengths arises from cool material outside the warm molecular core, in which the source of the luminosity is embedded.  The non-detection of any emission lines in the TReCS spectrum suggests that the mid-infrared emission that emerges from the nucleus is a warm dust continuum with weak spectral structure.

\section*{Acknowledgments}

 This paper is based on observations obtained at the Gemini Observatory, which was operated by the Association of Universities for Research in Astronomy, Inc., under a cooperative agreement 
with the NSF on behalf of the Gemini partnership: the National Science Foundation (United 
States), the Science and Technology Facilities Council (United Kingdom), the 
National Research Council (Canada), CONICYT (Chile), the Australian Research Council (Australia), 
MinistŽrio da Cincia, Tecnologia e Inova‹o (Brazil) 
and Ministerio de Ciencia, Tecnolog'a e Innovaci—n Productiva (Argentina).  This work is based in part on archival data obtained with the Spitzer Space Telescope, which is operated by the Jet Propulsion Laboratory, California Institute of Technology under a contract with NASA. 

AA-H acknowledges support from the Universidad de Cantabria through the Augusto G. Linares programme and from the Spanish Plan Nacional grant AYA2012-31447.

{}

\label{lastpage}

\end{document}